# Prediction of Retained Capacity and EODV of Li-ion Batteries in LEO Spacecraft Batteries

S. Ramakrishnan, S. Venugopalan, A. Ebenezer Jeyakumar

**Abstract**—In resent years ANN is widely reported for modeling in different areas of science including electro chemistry. This includes modeling of different technological batteries such as lead acid battery, Nickel cadmium batteries etc. Lithium ion batteries are advance battery technology which satisfy most of the space mission requirements. Low earth orbit (LEO)space craft batteries undergo large number of charge discharge cycles ( about 25000 cycles) compared to other ground level or space applications. This study is indented to develop ANN model for about 25000 cycles, cycled under various temperature, Depth Of Discharge (DOD) settings with constant charge voltage limit to predict the retained capacity and End of Discharge Voltage (EODV). To extract firm conclusion and distinguish the capability of ANN method, the predicted values are compared with experimental result by statistical method and Bland Altman plot.

—— ——— ◆ ——— ——

## 1 INTRODUCTION

Battery comprises complex set of interacting physical and chemical processes, the purpose of which is the conversion of chemical energy into electrical energy. The processes are often strongly influenced by the battery environmental conditions and usage profile. In fact, due to the number and complexity of the processes taking place, and the inability to accurately describe them, makes it difficult to develop an accurate battery model[1].

The Low Earth Orbit (LEO) spacecrafts are used for a wide variety of remote sensing applications such as weather forecast, cartography, urban planning, environmental assessment, agriculture, forestry, ground water management etc. Batteries are required in LEO spacecrafts for support during eclipse period and to meet peak power demand during sunlit. In LEO the batteries have to undergo 15 charge discharge cycles in a day. Typically the orbital period of a LEO spacecraft comprises 65 minutes of sunlit and 35 minutes of eclipse duration. Batteries are charged during the sunlit and discharged during eclipse to meet the spacecraft power demand. Batteries may be discharged even in sunlit to meet the peak power demand. In LEO the batteries are expected to last for 5 to 8 years. Consequently, LEO spacecrafts are highly demanding with respect to battery performance requiring 5500 charge discharge cycles in a year. Requirement of larger number of cycles forces to restrict the DOD, generally, to less than 30%. Lower the DOD, temperature and end of charge voltage (EOCV) limit, longer is the cycle life. In LEO orbits the batteries are used continuously without any rest or open circuit duration.

Lithium-ion batteries are high energy density rechargeable batteries that offer increased energy density, long cycle life, resistance to launch vibration, high reliability, wide temperature range of operation, radiation resistance, high round trip efficiency with no memory effect. These features of lithium-ion batteries lead to its selection in spacecraft programs. But this lithium-ion cell performance and the lifespan are largely dependent on parameters such as operating temperature, depth of discharge (DOD), and also charge discharge rates Since any failure of the batteries will lead to failure of spacecraft, they should be highly reliable and are to be tested extensively to stringent specification requirements before being put into use. The retained capacity of below 40% and EODV of less than 2.5V are considered to be failure of battery[2][3].

Basically, mathematical models of physical systems are constructed to facilitate our understanding of mechanisms that lead to specific responses and to enable response predictions .This creates the need for prediction tools that provide users with useful information such as remaining working time ,available energy at every desired time of operation, etc. In this regard, the Artificial Neural Networks (ANNs), one of the most powerful modeling techniques could be explored as a possible tool to predict the charge discharge characteristics of rechargeable batteries. Because, ANNs play a vital role in analyzing and predicting the behavior of systems that cannot be described by any analytical equations[4][5].

In recent years, multivariate methods and ANNs are used to predict the capacity behavior of lead-acid batteries alone, wherein literature is replete with reports on the modeling and prediction of characteristics of lead-acid batteries . In this regard, ANNs have been used to predict capacity and power, gust effects on a grid-interactive wind energy conversion system with battery storage, cycle life and failure mechanism of lead-acid system. ANN has been reported to understand and to predict possibly the rechargeable lithium-ion cell charge–discharge characteristics . The estimation (interpolation)

————————————————

• S.Ramakrishnan Research Scholar Anna University , Chennai, India, Phone: 919841192930.
• S. Venugopalan Head, Battery Division , ISAC Bangalore, India, Phone: 918025083523.
• A. Ebenezer Jeyakumar Director, SREC, Coimbatore, India, ,Phone: 9198487282966.

capability of ANN's are assessed by odd cycle charge–discharge behavior for even cycle training data approach, and vise versa which is a commonly reported methodology in the literature [4][6].

On the other hand the ANN's prediction ability largely depends on closeness of the training data sets and the error target for which ANN has been trained. At the same time it is very important that very close training data sets and large number of training data sets leads to large training time and high training error. Interpolation using ANN has limitations when ANNs are used to model lithium batteries for LEO space applications where large number of cycles are involved with the other parameters such as temperature and DOD. In the LEO space applications mostly load is fixed and charge rate is also constant. So the variations due to charge and discharge rates are not considered. Increase in number of cycles increases the number of data sets. This minimises the number of data sets and the cycle interval of data for which we can achieve less network training time and low error rate. When considering changes in the other parameters such as temperature and DOD which also interplay with the charge discharge cycle to define the retained capacity and EODV, it becomes a complex situation to train the ANNs.

Experiments were performed under various DOD, temperature settings for 25000 charge discharge cycles. to find retained capacity and EODV. The obtained experimental result datasets have to be minimized to achieve best results using ANN. Multivariate analyses have been performed to find the effect of individual parameters such as temperature, DOD, with respect to charge discharge cycles on retained capacity and EODV. The observations from estimated model using Multivariate analyses have been in use to define the datasets to train the ANN. This approach is found to be more appropriate in training the ANN. This paper is probably the first in reporting model using ANN for large number of cycles with high accuracy.

Comparisons between experimental results with the predictions by the model are made using statistical method and Bland Altman plot to find the accuracy of the designed ANN model for lithium ion battery for LEO space applications. In statistical method Average absolute percentage error (AAPE), correlation co-efficient ( R-Squired value) between observed and predicted value are taken as the performance measures. Bland Altman plot is deemed simple both to do and to interpret. The X axis will represent the mean of the comparison values and the Y axis with the difference between the values. This work is aimed at development of a model and also to study the impact of temperature, DOD and cycles on retained capacity and EODV of lithium-ion rechargeable batteries.

## 2 CELL TEST SET-UP

Experiments were performed using Maccor series 4000 battery/cell test system. The cell test system is designed specifically for running multi-channel high speed tests on cells. The system is modular in design with 48 channels. The system allows a batch of cells to be tested independently for various parameter settings such as EOCV limit, temperature, DOD, etc. For example, to test the cycle life expectancy with respect to a specified parameter, the set-up allows a number of cells to be charged independently to a predetermined voltage level at a predetermined rate and discharged at a predetermined rate up to an end voltage at different temperatures. The system makes use of analogue control loops for smooth control of applied signals.

For the present problem retained capacity of lithium-ion cells have been tested with constant end of charge voltage of 4.1 V for temperatures range $10^\circ$ C to $30^\circ$ C and DOD of 10% to 30% for 25000 charge discharge cycles. as tabulated in Table 1.

**TABLE 1**
Experiment Datasets

| Temperature | Depth of Discharge ( DOD ) |
|---|---|
| 10 º C | 10 % |
| 10 º C | 20 % |
| 20 º C | 20 % |
| 10 º C | 30 % |
| 20 º C | 30 % |
| 30 º C | 30 % |

## 3 ANN SOFTWARE DEVELOPMENT

Model development using ANN involves two steps, one is learning with the experimental data sets for low error target which results in generation of weights and biases and the other is prediction for noisy data (the data's not used in training process) using the generated weights and biases. Even though there are a number of network architectures available, the most widely used network is multi-layered feed-forward network[8][9] trained with the back-propagation learning algorithm with sigmoidel activation function. The training software and prediction software are developed using "C" program to provide flexible platform for varying number of hidden layers, Number of neurons in each layer, and learning rate[4][10].

The developed software provides convenience of generation of weights and biases files for a specified error target. The generated weights and bias files is used for further reduction in error target. This reduces training time and is useful to achieve minimum error target without over fit or under fit. After achieving minimum error target the generated weights and biases file is used by the prediction software for interpolation by which the retained capacity and EODV can be estimated.

## 4 MULTIVARIATE ANALYSIS

Multivariate analysis is performed to find the effect of the individual parameters DOD, temperature and charge discharge cycle on retained capacity and EODV for LEO profile. The experimental data sets with the interval of 1000 charge discharge cycle for different DOD and Tempera-ture settings as mentioned in Tabile1 have been taken as input parameter and multivariate analysis has been performed. The multivariate estimated model for retained capacity is shown in equation.(1) and equation.(2) provide the EODV estimated model.

Estimated model for Retained Capacity:
RC = 110.29-0.7551*T-0.2977*DOD-0.0014*C        **(1)**

Estimated model for EODV:
EODV=4.3156-0.1297*T-0.0093*DOD-7.1705E-06*C    **(2)**

Where  RC – Retained Capacity in %
T – Temperature in °C
DOD -  Depth of Discharge in %
EODV- End of Discharge voltage in Volts.
C - Charge and discharge cycle number

It is interesting to see from the estimated model equation for retained capacity the effect of DOD is higher than the temperature followed by the number of cycles. Where as EODV reduces largely with temperature followed by DOD and cycle. The initial internal resistance is high for lower temperature but change in resistance is less in due course of cycling in higher temperature where the initial resistance is low but change in resistance is high in due course of cycling. This investigation has been taken as prelude to decide the data set to train the ANN model.

## 5  ANN ARCHITECTURE FOR BATTERY MODEL

Two feed forward error back propagation networks have been developed for the present problem .One, to estimate retained capacity and the other for EODV for charge voltage limit of 4.1V with various temperature, DOD and charge-discharge cycles. In this network input layer consists of three neurons which are presented with the parameters temperature, DOD, and charge-discharge cycles and the output layer consists of one neuron - retained capacity /EODV. Different configurations have been tested to find the number of hidden layers and number of neurons in each layer. Out of the tested configurations, the configuration with two hidden layers having five neurons each and two hidden layers having nine neurons showed good prediction . But two hidden layers having nine neurons each is able to train the network for lower error target than the other. So it is found that the configuration with two hidden layers having nine neurons each is suitable for the present problem. The final configuration of the network is three neurons in the input layer, two hidden layers with nine neurons each and one neuron in the output layer as shown is figure 1.Both the networks are trained with 156 data sets each to predict retained capacity and EODV respectively. The ANNs have been trained with  learning rate of 0.4 with sigmoidal activation function . The error target achieved is 0.7% and 0.2% for retained capacity and EODV prediction models respectively.

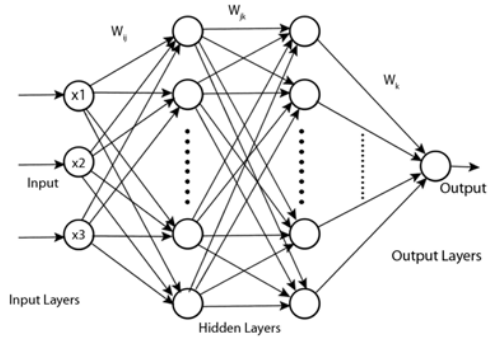

Fig 1 Back propagation neural network.

## 6  RESULTS AND DISCUSSION

Retained capacity estimation result shown in figure.2 depicts clearly the effect of DOD and temperature over charge discharge cycle. Retained capacity has been expressed in terms of percentage. It is observed from the plot that for the same temperature when the DOD increases the loss in capacity is increased. For example for the temperature of 10° C with the DOD of 10%, 20% and 30% the loss in the capacity is found to be 20%,35% and 45% respectively after 25000 cycles. At the same time for same DOD of 30% with different temperatures of 10°C, 20°C and 30°C the capacity fading is observed as 45%,52% and 60%. This indicates the influence of DOD in capacity fading of lithium ion battery is higher with the higher operating temperature. The highest capacity loss of 60% has been seen at 30°C with 30% DOD and lowest was observed at 10°C with 10% DOD after 25000 cycles. This indicates that for lower the DOD and lower temperature the capacity fading is also lower. It is interesting to note from the graph that the initial loss of capacity is more (observed from 5000 cycles) than subsequent capacity where the fading is observed to be uniform in nature.

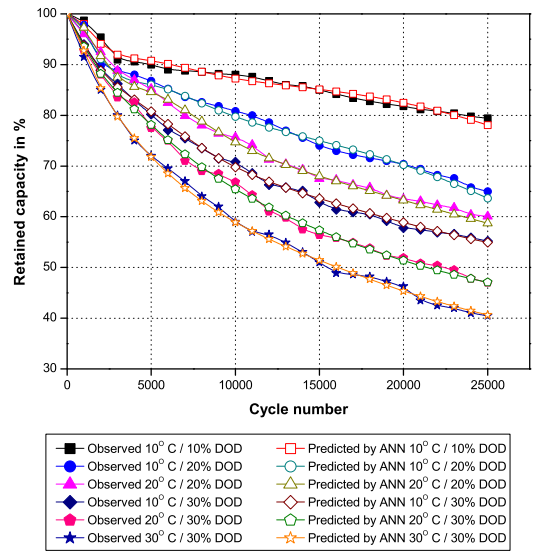

Fig. 2. Observed and Predicted values of Retained Capacity by ANN Model.

EODV graph shown in figure 3 demonstrates that with lower temperature initial internal resistance is high. So the drop is also high but in the case of higher temperature the initial internal resistance is low. So the drop is also less. But over the cycles it is seen that the internal resistance increases for high temperature over lower temperature. For 10°C with 10% DOD the loss in voltage after 25000 cycles is 0.04V and for 30°c with 30% DOD it is observed to be 0.35V.

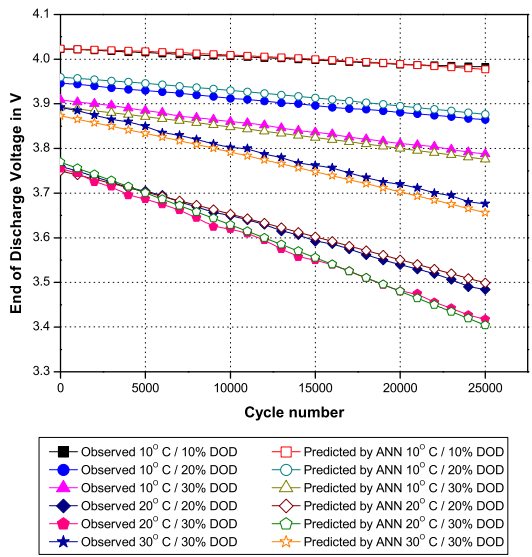

Fig. 3. Observed and Predicted values of EODV by ANN Model.

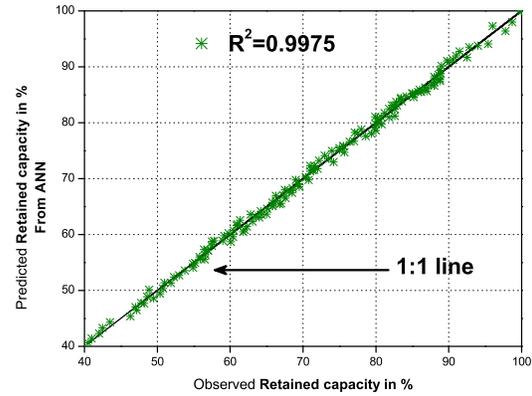

Fig. 4. Comparison graph between Observed and Predicted values of Retained Capacity using ANN.

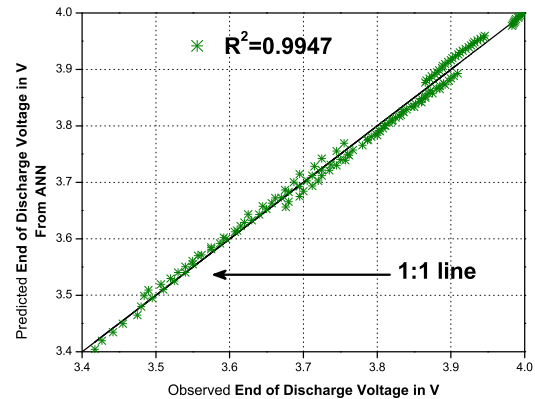

Fig. 5. Comparison graph between Observed and Predicted values of EODV using ANN.

Observed and predicted values of retained capacity and EODV plotted in the figures2 and 3 show an excellent agreement between them. The training has been done step by step reducing the training error rate. The verification has been done with test data sets and also with training data sets to avoid the danger of over fitting or under fitting.

The correlation coefficient for the relationship between the retained capacity predicted with the experiment result is 0.9975. The Average absolute percentage errors (AAPE) and coefficient of variation for the variables are 0.899 % and 0.0105 respectively.

The correlation coefficient for the relationship between the EODV predicted and the experiment result is 0.9947. The Average absolute percentage errors (AAPE) and coefficient of variation for the variables are 0.271 % and 0.0031 respectively.

The retained capacity and EODV predicted by the ANN models are compared with the experimental results to extract degree of prediction accuracy and generalization capability of the ANN. The retained capacity prediction and EODV prediction by the trained ANNs is plotted against the 1:1 line which represents 100% accuracy, as shown in figures 4 and 5. Visual inspection reveals that both the predictions are found to be very close with 1:1line.

Bland Altman plot comparison of observed experimental data with the predictions by ANNs for retained capacity and EODV is plotted in figures 6 and 7. It can be seen that variation between the observed and predicted retained capacities is negligible. In almost all cases the variation is within ±2%. For EODV the variation is -0.75% to +1.25%. This confirms that our models using ANN are very appropriate.

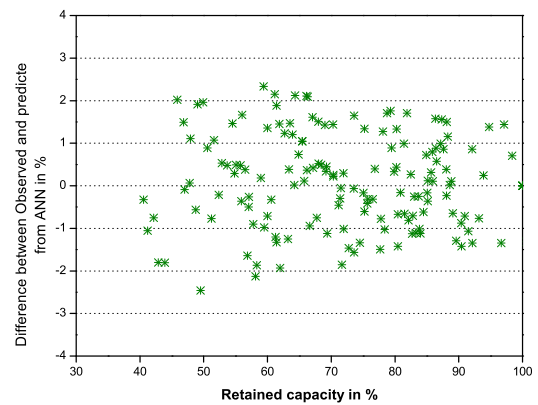

Fig. 6. Bland Altman Plot for Observed and Predicted values of Retained Capacity using ANN model.

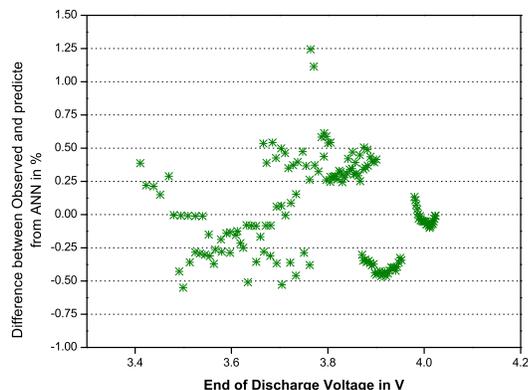

Fig. 7. Bland Altman Plot for Observed and Predicted values of End of Discharge Voltage using ANN model.

## 7 CONCLUSION

This paper demonstrated the development of ANN models for lithium-ion battery to predict the retained capacity and EODV where a large number of cycles is involved. The effect of individual parameters on retained capacity and EODV observed from the Multivariate estimated model is more useful to define the interval of datasets between the different input parameters, which inturn reduces the data sets required for training ultimately results in reduction in training time and also possibility to train the network to low error rate. Considering the complexity of relation between the input and the out put result obtained with the chosen data sets interval presented in this paper show encouraging results for prediction of the retained capacity and EODV. This proves the capability of the ANN to model non-linear complex systems. From the above study it is well established that the DOD and temperature over large charge discharge cycles influence the retained capacity and EODV of lithium-ion batteries.

**Professor S Ramakrishnan** is Master of Science (By research) from Anna University in the field of Electrical Engineering. Currently he is pursuing his research towards PhD from Anna University. He has had over 12 years of teaching experience after 5 years experience in the Industry. Currently he is Head, Department of IT in Sakthi Mariamman Engineering College, Chennai, India. He is a member of IEEE and IETE. He has presented papers in National and International conferences.

**Professor A. Ebenezar JeyaKumar** has graduated in Electrical Engineering from Annamalai University, Chidambaram, India. He has done Masters and PhD in High Voltage Engineering from Anna University, Chennai, India. He retired as Principal, Government College of Engineering, Salem, India. He has over 35 years of teaching experience in under graduate and post graduate programs. He was a Syndicate member of Anna University. He has published many papers in peer referred international journals and conferences. He has also authored books.

**Dr. S. Venugopalan** received his M.Sc (Chemistry) from I.I.T. Bombay in 1979 and Ph.D (Electrochemistry) from I.I.Sc, Bangalore in 1992. He served first as a Development Engineer and later as a Production and Quality Control Manager in M/s Tamilnadu Alkaline Batteries Madras during 1974 to 1982. He joined ISRO Satellite Centre (ISAC), Bangalore in 1982. Since then he is actively involved in the design and fabrication of batteries for Indian National Satellite Programs. Presently he is heading the Battery Division, Power Systems Group, at ISAC. His interest is in the field of electrochemical energy conversion and storage devices especially for aerospace applications. He has several technical publications and reports to his credit. He is an active member of the Electrochemical Society of India and Society for the Advancement of Electrochemical Science and Technology.